# Electrochemically Modeling a Non-Electrochemical System: Hydrogen Peroxide Direct Synthesis on Palladium Catalysts

*Min-Cheol Kim and Sang Soo Han\**

Computational Science Research Center, Korea Institute of Science and Technology, Seoul 02792, Republic of Korea

\*Corresponding Author: sangsoo@kist.re.kr (SSH)

# Abstract

Nonelectrochemical hydrogen peroxide direct synthesis (HPDS) under ambient conditions is an environmentally benign and energy-efficient process that produces a green oxidizer. Despite its industrial importance, the reaction mechanism of HPDS is still controversial, even for the prototypical catalyst Pd. Density functional theory (DFT) calculations with a comprehensive consideration of entropic and solvation effects reveal that the conventionally accepted Langmuir-Hinshelwood mechanism fails to explain why $H_2O_2$ production dominates over $H_2O$ production, which was experimentally reported. Inspired by the recently suggested heterolytic mechanism that involves electron and proton transfer at Pd catalysts, we propose a new electrochemical DFT model that is applicable for nonelectrochemical systems where a protonation intrinsically occurs. Our model is based on combining the Butler-Volmer equation and constant potential DFT with hybrid explicit-implicit solvent treatments. Application of this model to Pd(111) surfaces produces accurate descriptions of the activation barriers of both $H_2O_2$ and $H_2O$ production (within only ~0.1 eV of experimentally measured values). The heterolytic mechanism has a lower barrier for the protonation steps for $H_2O_2$ production than the nonelectrochemical hydrogenation steps, leading to advantageous kinetics for $H_2O_2$ production over $H_2O$ production. This work is the first theoretical and computational study supporting the heterolytic $H_2O_2$ production mechanism, and it resolves the unanswered discrepancies between previous experimental and DFT results. We expect that these results will readily help the systematic development of improved catalysts for $H_2O_2$ synthesis.

**Table of Contents**

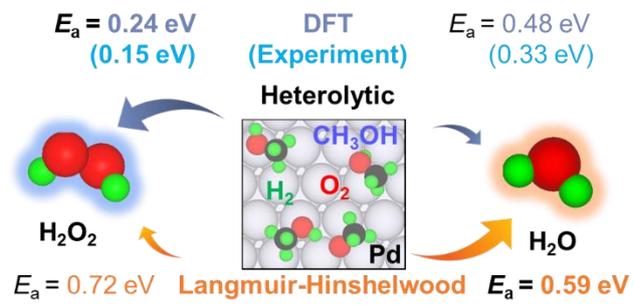

# 1. INTRODUCTION

Hydrogen peroxide ($H_2O_2$) is a well-known green oxidizer that is intensively used for disinfection and bleaching in the pulp and paper industries.[1] Unfortunately, the manufacturing process of $H_2O_2$ is not so "green", as it uses toxic anthraquinone and generates excessive amounts of greenhouse gases. Moreover, it is energy intensive and requires large-scale facilities.[2-3] To overcome such issues, hydrogen peroxide direct synthesis (HPDS) from $H_2$ and $O_2$ under ambient conditions has been regarded as a promising alternative.[4-6]

Most catalysts reported to perform HPDS under ambient conditions have been limited to Pd and its bimetallic derivatives such as Pd-Ru, Pd-Zn, Pd-Au, Pd-Sn, and Pd-Ni.[7-14] Despite many efforts in researching HPDS so far, the reaction mechanism of HPDS on metal catalysts is still controversial, even for the most prototypic Pd catalysts.[15] The generally accepted mechanism for the HPDS is the Langmuir-Hinshelwood (LH) mechanism (Figure 1a),[16-26] where surface-activated *H species (* denotes a surface site) react with *$O_2$ species adsorbed on the catalyst surface one-by-one to form $H_2O_2$. In particular, in first-principles studies, the LH mechanism has widely been accepted mainly due to its straightforwardness for modeling the catalytic reaction on metal surfaces. On the basis of the LH mechanism, several first-principles studies have explored the effects of coadsorbates such as –H,[17] –O,[19-20] and –Br;[21] various facets;[27] charge transfers by metal doping;[23-24] and surface oxidation.[18] On the other hand, Wilson and Flaherty experimentally suggested a heterolytic mechanism for HPDS on Pd nanoparticles (NPs) (Figure 1b),[28] where the feasibility of the LH mechanism was questioned based on the results that the kinetics of the HPDS reaction were inconsistent with those of the LH mechanism.

A first-principles approach using density functional theory (DFT) is very useful in unveiling reaction mechanisms at the atomic level. However, direct DFT investigations of the heterolytic mechanism are very limited, although one study claims that the heterolytic mechanism exhibits slow $H_2O_2$ production, indirectly treating the heterolytic mechanism via barrierless proton transfer reactions and concluding that HPDS proceeds via the LH mechanism.[19] Moreover, recent studies showed that the solvent plays an important role in the HPDS reaction,[28,29] although it might not be regarded as an important factor in the LH mechanism at first glance. To our knowledge, only a single DFT study considered the effect of solvents on HPDS, which it did by putting a single explicit water molecule within an implicit water solvent on small nanoclusters of ~30 atoms.[21] Additionally, most previous DFT studies used the reaction enthalpy rather than the Gibbs free energy when investigating the reaction pathway, although there are some exceptional cases.[24-25,30] In this regard, a systematic consideration of the solvation effects, electron-proton transfer, and free energy correction is required to clarify the HPDS reaction mechanism via DFT calculations.

Here, using DFT methods, we first investigate the feasibility of the LH mechanism for HPDS over Pd and find that although the consideration of solvation effects and a free energy correction can further improve the feasibility of the LH mechanism in comparison to that in previous DFT reports, the LH mechanism still has unsolved issues, such as the calculated rate-determining-step (RDS) barrier for the main reaction of HPDS ($H_2O_2$ production) being significantly higher than the experimental reaction barrier and the main reaction having a higher barrier than the side reaction ($H_2O$ production), indicating a sharp contrast to experimental observations. Therefore, we introduce a new model to capture the kinetics of the heterolytic mechanism of HPDS by applying the Butler-Volmer equation in conjunction with a constant-potential DFT method and a hybrid implicit-explicit solvent treatment. Within this model, the reaction barriers are found to be substantially lower than those of the LH mechanism for Pd and much closer to experimentally measured values (only ~0.1 eV difference). The new energetics provide adequate explanations for the $H_2O_2$ activity dominating $H_2O$ production, resolving the discrepancies between experimental observations and results from conventional theoretical models. This work is the first theoretical and computational study that provides supportive evidence for the heterolytic mechanism.

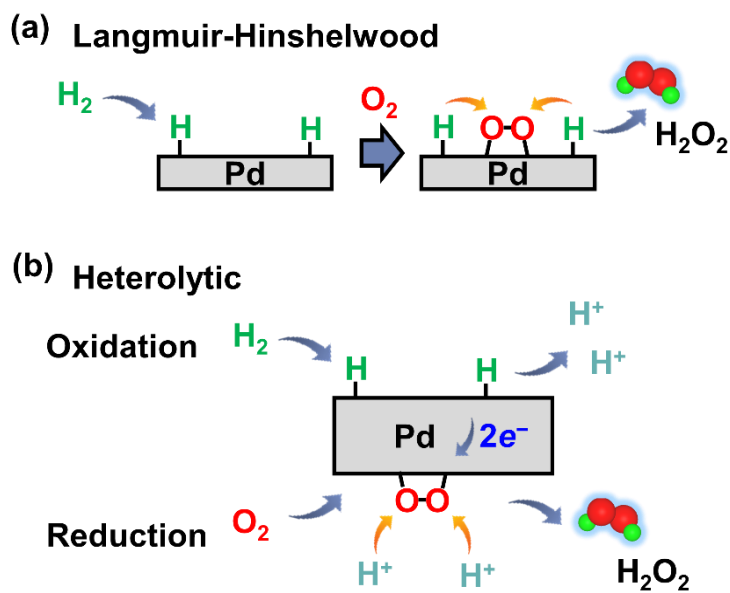

**Figure. 1.** Schematic diagrams for (a) the Langmuir-Hinshelwood (LH) mechanism and (b) the heterolytic mechanism of HPDS on Pd

## 2. COMPUTATIONAL DETAILS

All DFT calculations were performed using the Vienna *Ab Initio* Simulation Package (VASP)[31] in conjunction with VASPsol,[32] in which the revised Perdew–Burke–Ernzerhof (RPBE)[33] was used as the exchange and correlation functional. Grimme's D3 dispersion correction[34] was also considered because D3 correction has been known to greatly improve the accuracies for the energetics and structural information for metal-adsorbate[35] and metal-solvent interactions,[36] both of which are essential for this study. The projector-augmented-wave method was adopted to describe the ionic core potential,[37] and an energy cutoff of 500 eV was used. Monkhorst-Pack *k*-point meshes of 3×3×1 were used for all systems, where a vacuum spacing of 15 Å was used to prevent interslab interactions, and spin polarization and dipole correction were considered. It is well known that the (111) surface is the most stable facet for face-centered cubic (FCC) structures and occupies over 72 % of a typical Pd NP model surface according to the Wulff construction rule;[19] thus, we investigated all reactions on a Pd(111) slab. Each (111) surface was modeled with a supercell slab that consisted of a 4×4 surface unit cell with four layers of Pd. While the bottom two layers of Pd were fixed, the top two layers and adsorbate atoms were optimized until the energy change was less than $1.0 \times 10^{-4}$ eV cell$^{-1}$ and the force on each atom was less than 0.03 eV Å$^{-1}$ and 0.05 eV Å$^{-1}$ for vacuum and explicit solvent calculations, respectively.

For free energy corrections, corrections of enthalpy and entropy were estimated using vibrational frequencies of adsorbates evaluated using the finite difference method. The enthalpy correction H$_{corr}$ is:

$$\mathrm{H_{corr}} = E_{ZPE} + \int C_P dT \qquad (1)$$

where

$$\mathrm{E_{ZPE}} = \sum \frac{1}{2} \hbar \omega_i \qquad (2)$$

and

$$\int C_P dT \approx \int C_V dT = \sum \frac{\epsilon_i}{e^{\epsilon_i/k_B T} - 1} \qquad (3)$$

where $\omega_i$ is the vibrational frequency, and $\epsilon_i = \hbar \omega_i$.

The entropy of adsorbates was estimated assuming the harmonic oscillator limit as:

$$S = k_B \sum \left[ \frac{\epsilon_i}{e^{\epsilon_i/k_B T} - 1} - \ln\left(1 - e^{-\epsilon_i/k_B T}\right) \right] \quad (4)$$

The climbing-image nudged elastic band (CI-NEB) method was employed for transition state (TS) calculations. For the implicit solvent calculations, the linearized Poisson-Boltzmann (LPB) approach was used with ε = 33.1, which corresponds to the dielectric constant of methanol (methyl alcohol, MA) at 298.15 K,[38] with a Debye length of 3 Å, and a surface tension parameter τ of 0. As mentioned beforehand, we used the RPBE-D3 functional in this work. However, the parameters used in the LPB model were fitted with the GGA functional with no dispersion correction, which may raise a transferability issue.[39] Thus, we compared the solvation energies between RPBE and RPBE-D3 for $O_2$ adsorbed on the Pd(111) slab solvated with explicit MA molecules and found that the difference was marginal, being < 1 meV, justifying the use of RPBE-D3. To determine the initial structure for explicit MA solvent molecules, we performed an *ab initio* molecular dynamics simulation for 5 ps (picoseconds) at 298.15 K on the Pd(111) slab model with 20 Å thickness of β-phase solid MA[40] filled inside the simulation cell. After equilibration, we took a snapshot from the trajectory and removed all MA molecules except the lowest layer of 4 MA molecules. We also confirmed the feasibility of the explicit solvent model by removing the MA molecules one-by-one and whether the solvation energy was negative for each step (Figure S1).

## 3. RESULTS AND DISCUSSION

**3.1 HPDS via the LH Mechanism.** The most conventional way to investigate the LH reaction mechanism of HPDS on a metal surface is to calculate the adsorption energy ($E_{ads}$) of an intermediate of each reaction on the metal surface. Here, $E_{ads}$ is calculated as:

$$E_{ads} = E_{tot} - E_{slab} - E_{O_2(g)} - E_{H_2(g)}, \tag{5}$$

where $E_{tot}$ is the total energy of the slab with the adsorbate, i.e., the reaction intermediate; $E_{slab}$ is the energy of the slab only; and $E_{O_2(g)}$ and $E_{H_2(g)}$ are the energies of $O_2$ and $H_2$ gas molecules, respectively. Figure 2b shows the LH reaction pathway on the Pd(111) surface based on $E_{ads}$. This pathway is similar to those previously reported,[17, 19, 24] yet it raises several issues because Pd is a prototypic catalyst for HPDS.:

1. The *$H_2O_2$ detachment process (III →IV in Figure 2b) is endothermic by 0.51 eV.

2. The *$H_2O_2$ degradation step (III →VII in Figure 2b) has a very low energy barrier (0.07 eV), which is significantly lower (by 0.44 eV) than the *$H_2O_2$ detachment energy. In conjunction with issue 1, this indicates that *$H_2O_2$ would prefer to dissociate into *OH + *OH rather than be released from the Pd surface, implying that only a negligible amount of $H_2O_2$ would be produced.

3. The energy barrier at the RDS of the HPDS (0.78 eV in Figure 2b) is much higher than the experimental value (0.15 ± 0.02 eV).[28]

4. The energy barrier at the RDS of $H_2O_2$ production is higher than the RDS of $H_2O$ production, indicating a trend opposite of experimental observations.[28]

These facts suggest that Pd would not be a good catalyst for HPDS, in contrast to the experimental phenomena.

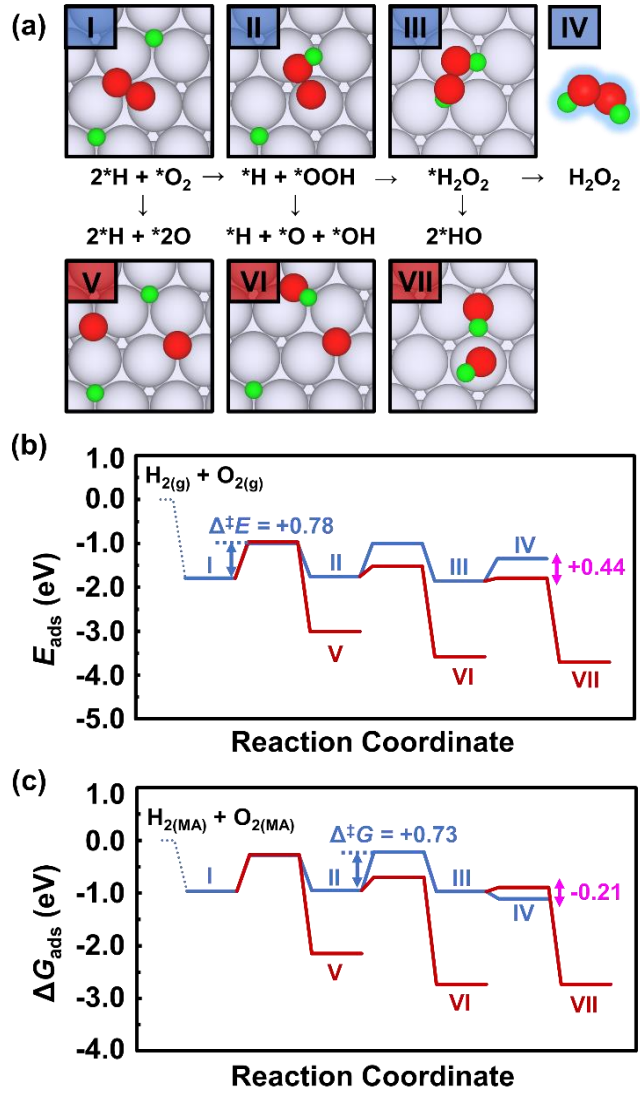

**Figure 2.** (a) HPDS reaction pathways for the LH mechanism, where the side reaction ($H_2O$ production) is also included. Color codes for the atoms are Pd = silver, O = red, and H = green. (b,c) Energetics for the HPDS reaction are represented by (b) $E_{ads}$ and (c) $\Delta G_{ads}$, following the pathway shown in (a). The energy barrier (blue number) for the RDS and the energy difference (pink number) between step IV and the TS of reaction III → VII are included.

To solve these issues, we considered the free energy ($G_{ads}$) instead of $E_{ads}$, as follows:

$$G_{ads} = G_{tot} - G_{slab} - G_{H_2(solv)} - G_{O_2(solv)} \tag{6}$$

where the free energy $G$ is defined as follows:

$$G = E + H_{\text{corr}} - TS + G_{\text{solv}} = E + E_{ZPE} + \int C_P dT - TS + G_{\text{solv}} \qquad (7)$$

Here, $E$ is the electronic energy, $H_{\text{corr}}$ is the enthalpy correction, $TS$ is the entropy correction, and $G_{\text{solv}}$ is the solvation energy. $H_{\text{corr}}$ and $TS$ are obtained from vibrational frequency calculations, and $G_{\text{solv}}$ is calculated using an implicit solvent model. Since MA is a commonly used solvent for HPDS,[28] we chose it as the solvent in this work. To represent the HPDS under ambient conditions, the temperature is set as $T = 298.15$ K.

Figure 2c shows the LH reaction energetics based on $G_{\text{ads}}$. The most notable features are that the *$H_2O_2$ detachment process (III → IV) is now exothermic and the $H_2O_{2(MA)}$ energy level (IV) is actually lower than the TS of the *$H_2O_2$ degradation reaction by 0.21 eV. This can be easily understood from the solvation and entropy contributions, as shown in Table S1. The *$H_2O_2$ (III) has fewer degrees of freedom in motion than $H_2O_{2(MA)}$ (IV), leading to a smaller entropy contribution for the *$H_2O_2$. Similarly, $H_2O_{2(MA)}$ is surrounded by more MA solvent molecules than *$H_2O_2$, and the solvation energy is stronger for $H_2O_{2(MA)}$ than for *$H_2O_2$; thus, the $H_2O_{2(MA)}$ is further stabilized compared to *$H_2O_2$. As a result, the $G_{\text{ads}}$ of $H_2O_{2(MA)}$ is lower than that of *$H_2O_2$ due to the contributions of solvation and entropy even though $H_2O_{2(MA)}$ has a higher $E_{\text{ads}}$.

Although one can solve the issue regarding the detachment of *$H_2O_2$ through consideration of the free energy correction and the solvation effect, other issues are still unsolved. The RDS energy barrier of HPDS was calculated to be 0.73 eV, which is much higher than the experimental value (0.15 ± 0.02 eV).[28] Because the $H_2O_2$ degradation barrier (0.08 eV) is still low compared to the RDS barrier for HPDS (0.73 eV), the side reaction ($H_2O$ production) would likely be preferred. However, it was reported that the experimental energy barrier for the side reaction ($H_2O$ production) on Pd is 0.33 ± 0.03 eV, which is higher than the HPDS barrier (0.15 ± 0.02 eV),[28] indicating that the HPDS reaction is preferred to the side reaction. To provide a more comprehensive understanding of the relative energetics between the main reaction and the side reaction, we carefully investigated all possible $H_2O$ production pathways on the Pd(111) surface (Figure S2), which are lacking in Figure 2c. Here, the evaluated RDS barrier of $H_2O$ production was 0.66 eV, which is lower than the RDS barrier for $H_2O_2$ production. This is still insufficient to explain the experimental phenomenon that the reaction barrier for $H_2O_2$ production is lower than that of $H_2O$ production (Table 1). In addition, we double-checked the full paths for $H_2O_2$ production and $H_2O$ production using explicit solvents. Implicit solvent models treat the solvent-solute interaction as an average electrostatic field of the solvent; thus, they might not capture the effects of the local electric field on the reaction kinetics.[39] Despite the implementation of explicit solvents in calculations for the LH mechanism, the RDS barrier for $H_2O_2$ production barely changed (Figure S2, Table 1). The RDS of $H_2O$ production is the $O_2$ dissociation step (I → V in Figure 2-a) with a RDS activation barrier of 0.59 eV, which is lower than that (0.72 eV) of $H_2O_2$ production (Table 1), diverging from the experimental trend. Consequently, we now approach the conclusion that there exists a limit in the LH mechanism for HPDS on Pd. Therefore, we shifted our attention to the heterolytic mechanism.

**Table 1.** The calculated reaction barriers at the RDSs for the main reaction ($H_2O_2$ production) and side reaction ($H_2O$ production) of HPDS based on the LH and heterolytic mechanisms. Implicit solvent (IS) and explicit solvent (ES) models are considered for the LH mechanism. For comparison, the reported experimental values from Ref. 28 are included.

| Barrier (eV) | Experiment[28] | DFT | | |
| --- | --- | --- | --- | --- |
| | | LH-IS | LH-ES | Heterolytic |
| $H_2O_2$ | 0.15 ± 0.02 | 0.73 | 0.72 | 0.24 |
| $H_2O$ | 0.33 ± 0.03 | 0.66 | 0.59 | 0.48 |

**3.2 Electrochemical Model for a Non-Electrochemical System.** To compare the kinetics of the heterolytic and LH mechanisms, it is essential to obtain the free energy barrier $\Delta G^{\ddagger}$ of each reaction pathway. For the heterolytic reaction on a metal surface, this requires the estimation of $\Delta G^{\ddagger}$ for electrochemical charge transfer and proton transfer reactions. The difficulty lies in the fact that most DFT methods used for calculating $\Delta G^{\ddagger}$ in electrochemical reactions evaluate $\Delta G^{\ddagger}$ for a given electrochemical potential $U$. Defining $U$ is difficult in the nonelectrochemical HPDS because there is no external electrochemical potential applied to the system. However, the heterolytic mechanism can be regarded as a spontaneous redox reaction occurring on a single Pd electrode where the redox reactions are:

$$(\text{Anode}) \quad H_2 \rightarrow 2H^+ + 2e^- \tag{8}$$

$$(\text{Cathode}) \quad O_2 + 2H^+ + 2e^- \rightarrow H_2O_2 \tag{9}$$

Here, our objective is to find $\Delta G^{\ddagger}$ and $U$ for a chemical equilibrium between the anodic and cathodic reactions.

The kinetics of a redox reaction occurring on the same electrode can be expressed with the Butler-Volmer equation[41] as:

$$j = j_a - j_c = j_0 \cdot \left\{\exp\left[\frac{\alpha_a nF}{RT}(U - U_{eq})\right] - \exp\left[\frac{\alpha_c nF}{RT}(U - U_{eq})\right]\right\} \quad (10)$$

where $j$ is the electrode current density; $j_a$ and $j_c$ are the anodic and cathodic current densities, respectively; $U_{eq}$ is the equilibrium potential; $T$ is the absolute temperature; $n$ is the number of electrons involved in the electrode reaction; $F$ is the Faraday constant; $R$ is the universal gas constant; and $\alpha_a$ and $\alpha_c$ are the anodic and cathodic charge transfer coefficients, respectively. For simplicity of equations used later on, the sign of $j_c$ is defined oppositely to its conventional usage. At equilibrium, $j = 0$ and $j_a = j_c$. Thus, finding $U$ at equilibrium is identical to finding $U$ when $j_a = j_c$. By definition, $j_a$ and $j_c$ are related to the reaction rate of the oxidation ($\nu_a$) and reduction ($\nu_c$) as follows:

$$\nu_a = k_a c_{ox} = j_a/nF \quad (11)$$

$$\nu_c = k_c c_{red} = j_c/nF \quad (12)$$

From the Arrhenius equation,

$$k_a = A_a \exp\left[-\Delta G_a^{\ddagger}/RT\right] \quad (13)$$

$$k_c = A_c \exp\left[-\Delta G_c^{\ddagger}/RT\right] \quad (14)$$

where $k$, $A$, and $\Delta G^{\ddagger}$ are the rate constant, pre-exponential factor, and free energy barrier, respectively. Because $A_a c_{ox} = A_c c_{red}$ from the collision theory,[42] our problem is reduced to finding $U$ when $\Delta G_a^\ddagger = \Delta G_c^\ddagger$. For a given $U$, $\Delta G^\ddagger$ can be calculated with a constant potential DFT method in terms of the change in grand canonical free energy $\Delta\Omega$.[39, 43-44] Here, the grand canonical free energy $\Omega$ can be obtained by incorporating the exchange of energy with an external electron reservoir:[44]

$$\Omega(U) = G(U) - qU = G(U) - CU(U - U_0) \tag{15}$$

where $q$ is the surface charge of the electrode, $U$ is the potential equivalent to the chemical potential of the electron reservoir, $U_0$ is the system potential at zero charge, and $C$ is the capacitance defined as follows:

$$\left.\frac{\partial U_0}{\partial q}\right|_{q=0} = \frac{1}{C} \tag{16}$$

$U$ can be regarded as a negative of the Fermi energy of the electron reservoir in a system. In obtaining $U$, we followed the method proposed by Gauthier et al., which can avoid calculations with viciously large vacuum levels.[44] Then, one can scan $U$ and $\Omega$ by changing the number of fractional electrons in the computational supercell until $\Delta\Omega_a^\ddagger = \Delta\Omega_c^\ddagger$, where $\Delta\Omega_a^\ddagger$ and $\Delta\Omega_c^\ddagger$ are the grand canonical free energy differences between the TSs and the reactants of the anodic and cathodic reactions, respectively. This procedure would be exhausting. Fortunately, we found that it can be approximated that $U$ and $\Delta\Omega_{(a \text{ or } c)}^\ddagger$ are linearly related to each other, and thus, $\Delta\Delta\Omega^\ddagger = \Delta\Omega_a^\ddagger - \Delta\Omega_c^\ddagger$ can be approximated to have a linear relationship with $U$ by expressing $j_a$ and $j_c$ with respect to $U$ and $\Delta\Omega^\ddagger$:

$$j_a = j_0 \cdot \left\{\exp\left[\frac{\alpha_a nF}{RT}(U - U_{eq})\right]\right\} = A_a \exp[B_a U]$$

$$= nFc_{ox}A_a \exp[-\Delta G_a^{\ddagger}/RT] = C_a \exp[D_a \Delta G_a^{\ddagger}], \tag{17}$$

$$j_c = j_0 \cdot \left\{\exp\left[\frac{\alpha_c nF}{RT}(U - U_{eq})\right]\right\} = A_c \exp[B_c U]$$

$$= nFc_{red}A_c \exp[-\Delta G_c^{\ddagger}/RT] = C_c \exp[D_c \Delta G_c^{\ddagger}], \tag{18}$$

where *A, B, C,* and *D* are constants in approximations of the Butler-Volmer equation. Surprisingly, Gauthier et al. reported that $\Delta \Omega^{\ddagger}$ is indeed linearly related to *U*, which was derived from a different theoretical background grounding than ours.[44]

In the heterolytic HPDS reaction, the redox reaction pairs can be regarded as follows:

$$1/2\ H_2 \rightarrow H^+ + e^- \ //\ O_2 + H^+ + e^- \rightarrow {}^*OOH \tag{19}$$

$$1/2\ H_2 \rightarrow H^+ + e^- \ //\ OOH + H^+ + e^- \rightarrow {}^*H_2O_2 \tag{20}$$

Since $H_2$ dissociation on the Pd (111) surface has a relatively small barrier of 0.20 eV (Figure S3), we approximated the barrier of the anodic reaction as the oxidation barrier of *H activated on the surface. Thus, the redox reaction pairs can be expressed again as follows:

$$^*H \rightarrow H^+ + e^- \ //\ O_2 + H^+ + e^- \rightarrow {}^*OOH \tag{21}$$

$$^*H \rightarrow H^+ + e^- \ //\ OOH + H^+ + e^- \rightarrow {}^*H_2O_2 \tag{22}$$

Directly modeling $H^+$ in a periodic system is tricky in constant-potential DFT because one has to start from a structure relaxation of a zero-charge supercell. This might be solved using a localized electron method such as DFT+U; however, using the same U parameter for all chemical reactions would be an ill-defined approach due to the differences in chemistry or physics for the reactions.[45] Therefore, we used a slight modification to an approximation suggested by Akhade et al.[46] The barrier ($\Delta\Omega^{\ddagger}$) for the following reaction:

$$*A + H^+ + e^- \rightarrow *AH \tag{23}$$

$\Delta\Omega^{\ddagger} = \Omega^{TS} - \Omega[*A + H^+ + e^-]$, can be approximated as the $\Delta\Omega^{\ddagger*}$ of the following reaction:

$$*A + *H \rightarrow *AH \tag{24}$$

where $\Delta\Omega^{\ddagger*} = \Omega^{TS*} - \Omega[*A + *H]$, and $\Omega^{TS*}$ is the free energy of the TS for the reaction in Eq. 24. Applying this approximation to Eqs. 21 and 22, the redox reaction pair for HPDS can be regarded as the follows:

$$*H + * \rightarrow * + *H \;//\; *O_2 + *H \rightarrow *OOH + * \tag{25}$$

$$*H + * \rightarrow * + *H \;//\; *OOH + *H \rightarrow *H_2O_2 + * \tag{26}$$

Here, explicit MA solvent molecules were considered proton acceptors for the anodic reaction, which is a reverse Volmer reaction ($*H \rightarrow * + H^+ + e^-$), and proton donors for the cathodic reaction, which is a Heyrovsky-like reaction ($A + H^+ + e^- \rightarrow *AH$) (Figure S4).
Now we can find $U$ and $\Delta\Omega^{\ddagger*}$ that satisfy the following conditions:

$$\Delta\Omega_a^{\ddagger*} - \Delta\Omega_c^{\ddagger*} = (\Omega_a^{TS*} - \Omega[*H + *]) - (\Omega_c^{TS*} - \Omega[*O_2 + *H]) = 0 \tag{27}$$

$$\Delta\Omega_a^{\ddagger*} - \Delta\Omega_c^{\ddagger*} = (\Omega_a^{TS*} - \Omega[*H + *]) - (\Omega_c^{TS*} - \Omega[*OOH + *H]) = 0 \tag{28}$$

**3.3 HPDS via the Heterolytic Mechanism.** Using the model discussed in section 3.2, we investigated the reaction pathway for the heterolytic mechanism (Figure 3). For the main reaction, $*O_2$ and $*OOH$ are protonated through the reactions in Eqs. 19 and 20 (Figure 4a). To calculate $\Delta\Omega^{\ddagger*}$ for the protonation steps, we first estimated the capacitance $C$ for each state (I and A; II and B in Figure 4a) involved in the protonation step using Eq. 16 (Figure S5). Then, using the estimated $C$, we plotted $\Omega$ vs. $U$ for each state and calculated $\Delta\Omega^{\ddagger*}$ with Eqs. 27 and 28 (Figure S6 and Figure S7). Our approach reveals that $\Delta\Omega^{\ddagger*}$ for the first and second protonation steps is 0.24 and 0.23 eV, respectively, which is not only significantly lower than the values predicted for the LH mechanism (0.66 and 0.73 eV with implicit solvent models, 0.72 and 0.54 eV with explicit solvent models) but is also even closer to the experimental reaction barrier for the HPDS reaction (Table 1). In the heterolytic mechanism, the proton involved in the reaction is likely weakly bound to solvent molecules, leading to a Heyrovsky-like reaction with a small reaction barrier.

We also explored the side reaction with the heterolytic mechanism. The reaction barriers for the dissociation steps of $*O_2$ and $*OOH$ (Figure S2c) are higher than those of the protonation steps in Figure 3b (I → V: 0.59 eV vs. I → II: 0.24 eV, and II → VI: 0.27 eV vs. II → III: 0.23 eV), and the next reaction steps for $H_2O$ formation show energy barriers of 0.39 and 0.50 eV, respectively (Figure 3c), which implies that $H_2O$ formation followed by the dissociation of $*O_2$ or $*OOH$ would be less favorable than $H_2O_2$ formation. On the other hand, $*H_2O_2$ degradation into $*OH + *OH$ shows a reaction barrier of 0.48 eV, and then the $H_2O$ formation from $*OH + *OH$ shows a marginal energy barrier of 0.12 eV, indicating that the $*H_2O_2$ degradation step is the RDS for $H_2O$ formation. As a result, the reaction barrier for $H_2O$ formation via the heterolytic mechanism is 0.48 eV, which is higher than that (0.24 eV) for $H_2O_2$ formation. Now, the RDS barriers for the main and side reactions of HPDS are more similar to the experimental values than those obtained by the LH mechanism (Table 1).

Finally, we discuss the possible improvements in our electrochemical DFT model to represent the heterolytic mechanism. A microkinetic study of HPDS using our precise energetics could lead to a more accurate alignment between experimental reaction barriers and RDS barriers and enable an analysis of coverage effects. Treating the electron-proton transfer steps with the microkinetic modeling itself is challenging;[47] however, combining this with nonelectrochemical reaction steps complicates the matter further. We used the Butler-Volmer equation to describe the kinetics of one-electrode redox reactions with the assumption that the charge transfer coefficient ($α$) is independent of $U$. Strictly speaking, $α$ is a function of $U$, although $\varDelta\varOmega^{\ddagger}$ does not deviate from its linear correlation with $U$ in this work as shown in Figure S6. However, it is possible that for some cases one would need to evaluate $α(U)$ for both the anodic and cathodic reactions to obtain a more accurate reaction barrier. In addition, in calculating $\varDelta\varOmega^{\ddagger}$ with a constant-potential DFT method, it is assumed that the ionic motion adiabatically follows the atomic motion associated with the electrochemical reactions. The physical movement of atoms during a proton transfer, however, is on the order of fs (femtoseconds), while $H^+$ ion rearrangement is in the few-ps range.[42,48] Thus, a further investigation is required to address the effect of the discrepancy in the timescales. As an example, calculating $\Delta G^{\ddagger}$ with the Marcus theory while using constrained DFT to estimate the coupling and reorganization energies[49-50] might provide more accurate reaction barriers.

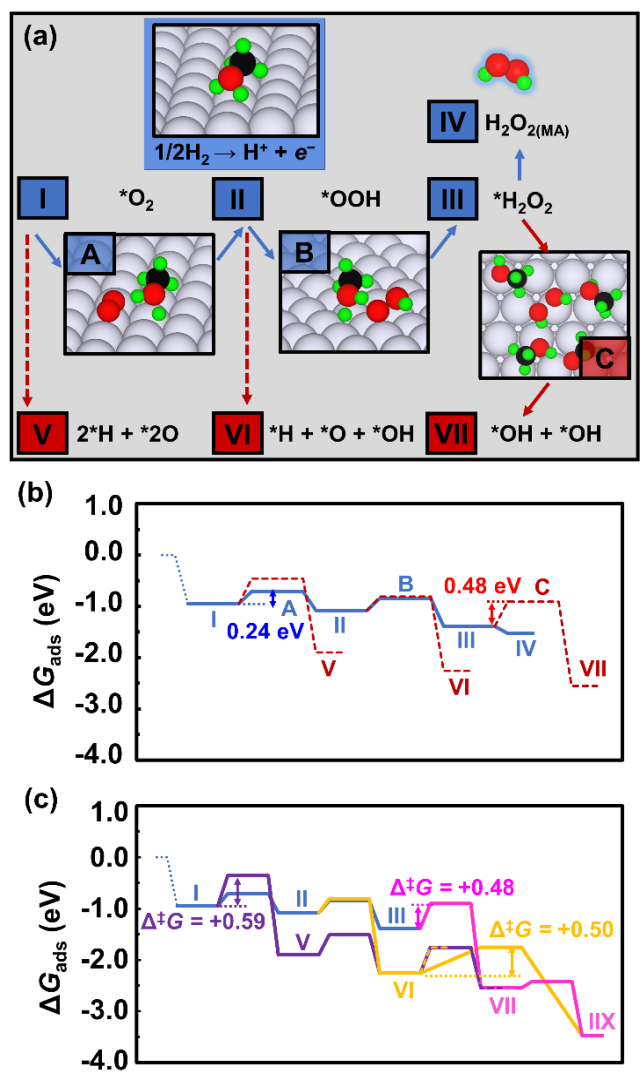

**Figure 3.** (a) Reaction pathways and TS structures and (b) energetics for the HPDS reaction and (c) the side reactions under the heterolytic mechanism. In (a), the color codes for Pd, O and H are silver, red and green, respectively. The energy barriers at the RDSs of HPDS and $H_2O$ production are depicted in (b) and (c). The dashed red line in (b) is the free energy change for the first few steps of the side reaction.

## 4. CONCLUSION

Although a more comprehensive model including entropic and solvent treatments can resolve a few issues found in previous DFT calculations for the HPDS reaction on Pd based on the LH mechanism, the conventional LH mechanism fails to capture the thermodynamics and kinetics of the HPDS reaction and its side reaction. Inspired by the heterolytic mechanism suggested by Wilson et al.,[28] a new DFT model was proposed for calculations of reaction free energies based on the Butler-Volmer equation, constant-potential DFT and a hybrid implicit-explicit solvent model. Using this approach, we revisited the main and side reactions of HPDS on the Pd (111) surface and found that the heterolytic mechanism is more plausible on Pd than the LH mechanism. To accurately understand the HPDS reaction over metallic surfaces within the DFT framework, it is important to consider a more comprehensive model that includes the heterolytic mechanism, free energy corrections and explicit solvation. Application of this new model to other HPDS catalysts will immediately help to understand how the catalysts work and how they should be improved. Consequently, this model can assist in the systematic design of efficient catalysts for HPDS with high selectivity under ambient conditions.

## ASSOCIATED CONTENT

**Supporting Information**

Details on additional results and discussions (PDF).

## AUTHOR INFORMATION

**Corresponding Authors**

*E-mail: sangsoo@kist.re.kr (S.S.H.).21

**Notes**

The authors declare no competing financial interest

## ACKNOWLEDGMENTS

We thank Dr. Donghun Kim and Dr. Hyun S. Park in KIST for their fruitful discussions. This work was supported by the Creative Materials Discovery Program through the National Research Foundation of Korea (NRF-2016M3D1A1021141).

Supporting Information

# Electrochemically Modeling a Non-Electrochemical System: Hydrogen Peroxide Direct Synthesis on Palladium Catalysts

Min-Cheol Kim and Sang Soo Han*

Computational Science Research Center, Korea Institute of Science and Technology, Seoul 02792, Republic of Korea

*Corresponding Author: sangsoo@kist.re.kr (SSH)

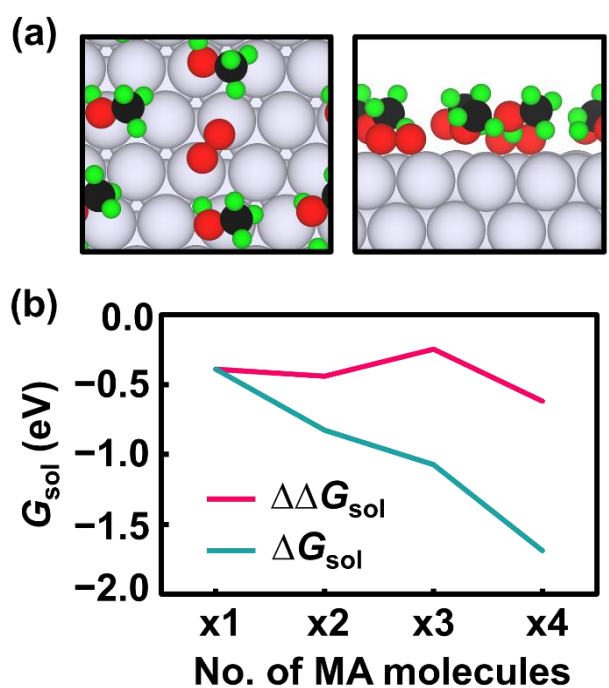

**Figure S1.** (a) The top (left) and side (right) views for the explicit methanol (methyl alcohol, MA) solvent model with *$O_2$ adsorbate on the Pd surface. (b) Solvation free energy vs. the number of solvent molecules in the simulation supercell depicted in (a). $\Delta G_{sol} = G[\text{slab} + \text{MA}] - G[\text{slab}]$ and $\Delta\Delta G_{sol} = \Delta G_{sol}[N] - \Delta G_{sol}[N-1]$, where $\Delta G_{sol}[N]$ is $\Delta G_{sol}$ for the supercell with N explicit MA molecules in the system. $\Delta\Delta G_{sol}$ is confirmed to be negative at N values up to 4. Color codes for atoms are Pd = silver, O = red, C = black, and H = green.

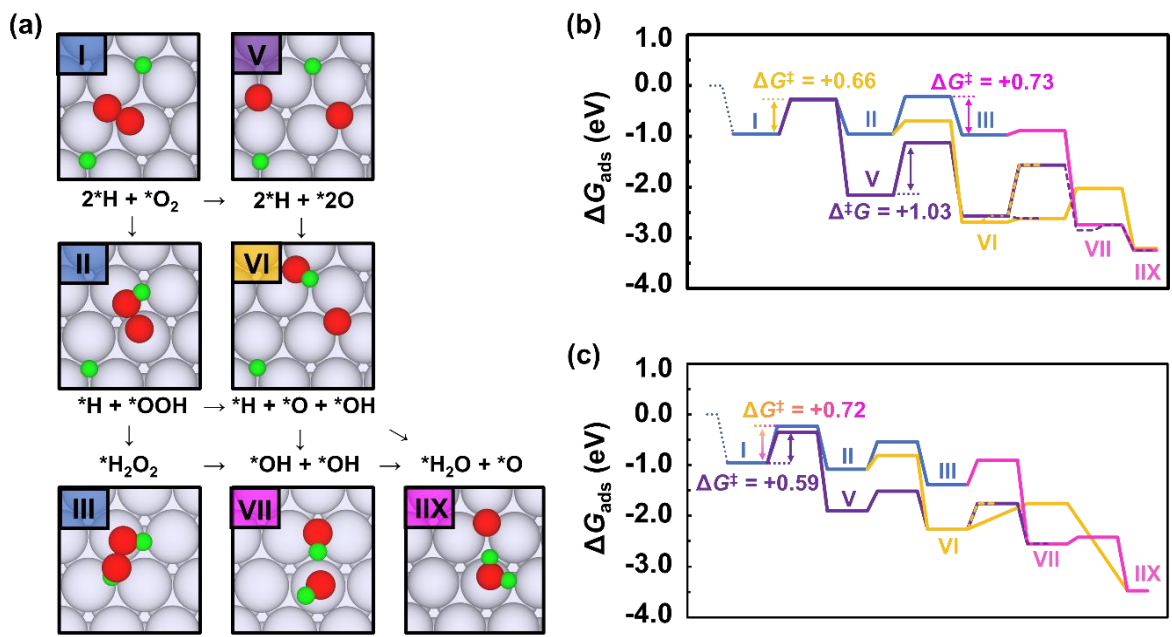

**Figure S2.** (a) Reaction pathways and (b, c) their energetics for the side reaction ($H_2O$ production) of HPDS in the LH mechanism using an (b) implicit solvent and (c) explicit solvent model. In (a), the color codes for Pd, O and H atoms are silver, red and green, respectively. In (b) and (c), the energy barriers for the RDS are included for each pathway.

We further investigated the side reaction of HPDS, i.e., $H_2O$ production, using the free energy LH model. Here, we considered the $H_2O$ production pathways involving the dissociation of *OOH (Eqs. 1 and 2) and *$H_2O_2$ (Eqs. 3 and 4):

$$*H + *OOH \rightarrow *H + *OH + *O \quad (1)$$
$$*H + *OH + *O \rightarrow *H_2O + *O \quad (2)$$
$$*H_2O_2 \rightarrow *OH + *OH \quad (3)$$
$$*OH + *OH \rightarrow *H_2O + *O \quad (4)$$

In addition, we considered the alternative $H_2O$ production pathways involving *$O_2$ dissociation (Eqs. 5 and 6) and the hydrogenation of *O (Eq. 7):

$$2*H + *O_2 \rightarrow 2*H + 2*O \quad (5)$$
$$2*H + 2*O \rightarrow *H + *OH + *O \quad (6)$$
$$*H + *OH + *O \rightarrow *OH + *OH \quad (7)$$

Figure S2 shows the $H_2O$ production pathways and their energetics: *OOH dissociation path (yellow), *$O_2$ dissociation path (purple), and *$H_2O_2$ dissociation path (purple). Here, $H_2O$ production following the *OOH dissociation path has the lowest RDS barrier (0.66 eV), while the *$O_2$ dissociation and *$H_2O_2$ dissociation paths have RDS barriers of 1.03 eV and 0.73 eV, respectively. From these results, the RDS barrier of $H_2O$ production is lower than that of $H_2O_2$ production.

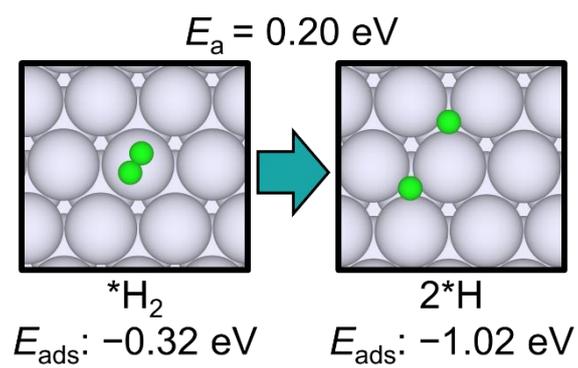

**Figure S3.** Atomic structures and energetics for $H_2$ dissociation on a Pd (111) surface. The calculated reaction barrier is 0.20 eV.

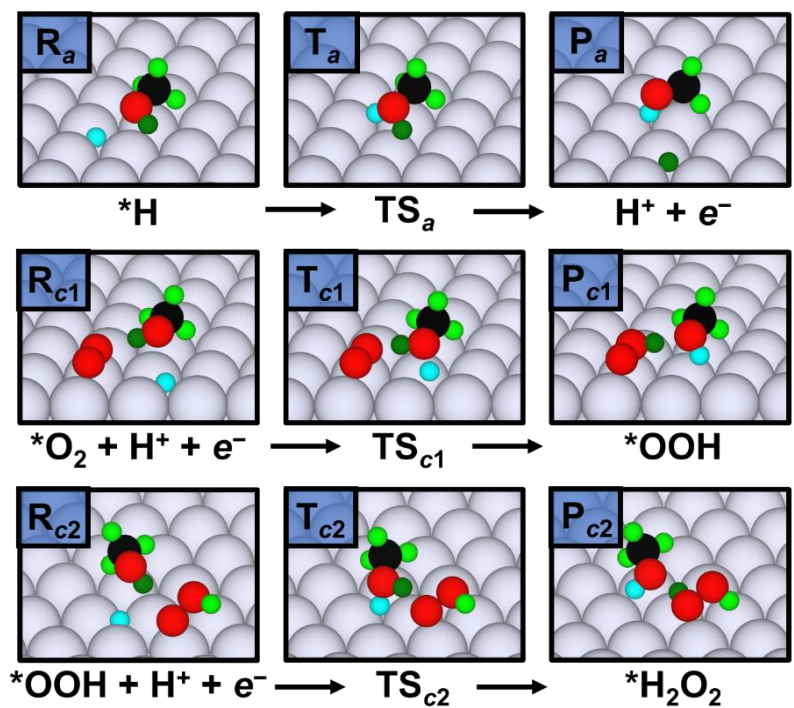

**Figure S4.** The reactant (R), transition state (T), and product (P) structures for the protonation steps of the heterolytic mechanism. The subscripts a, c1, and c2 denote the anodic reaction and the first and second cathodic reactions, respectively. MA molecules that do not participate in the reaction are removed for clarity. Color codes for atoms are Pd = silver, O = red, C = black and H = green. The protons involved in the reaction are indicated in cyan and dark green.

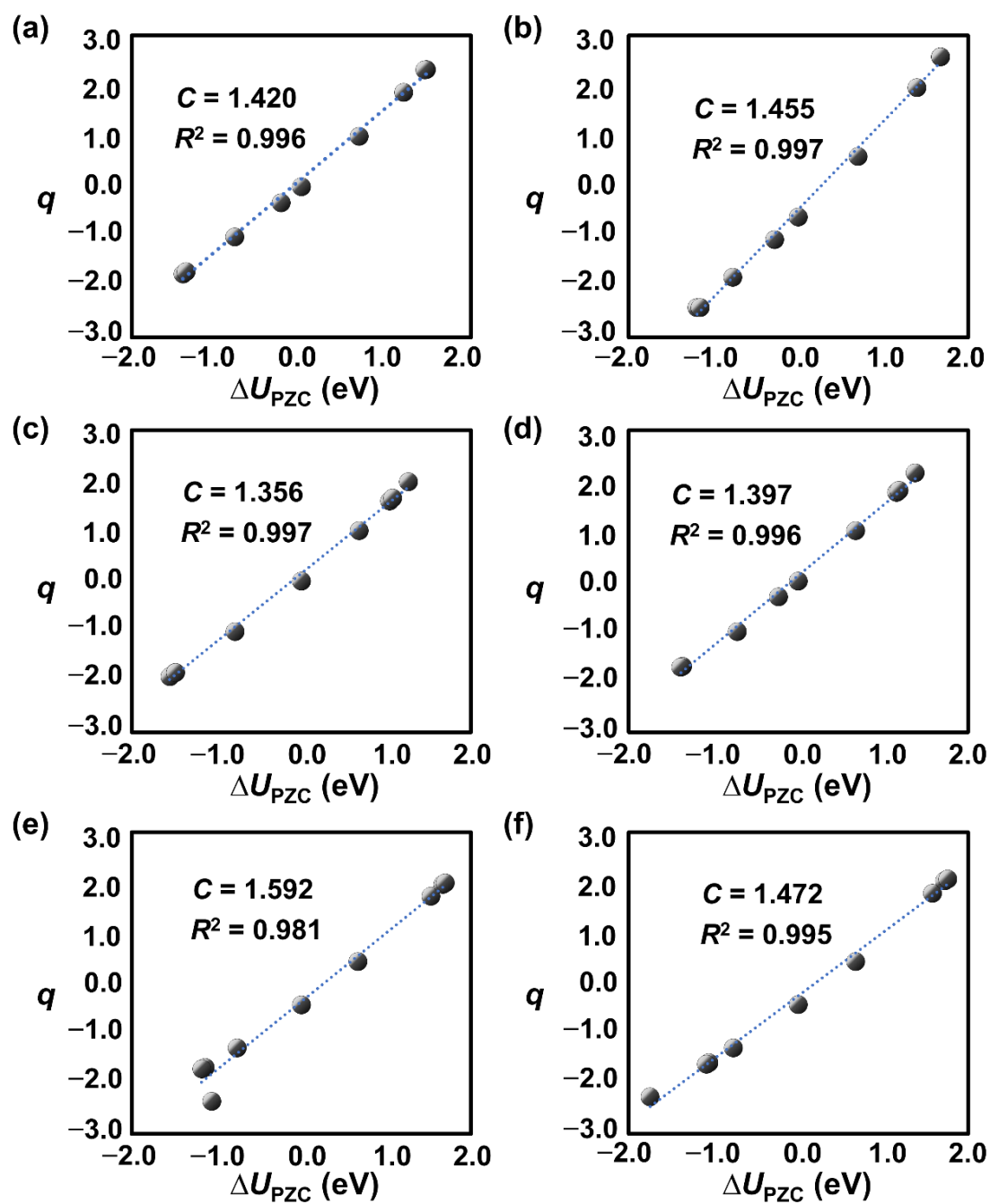

**Figure S5.** Charge of the system, $q$, plotted against the corresponding $\Delta U_{PZC}$, the system potential relative to the potential of zero charge (PZC), for (a) $R_a$, (b) $T_a$, (c) $R_{c1}$, (d) $T_{c1}$, (e) $R_{c2}$, and (f) $T_{c2}$ as in Fig. S2. The capacitance $C$ and the $R^2$ value are depicted for each state.

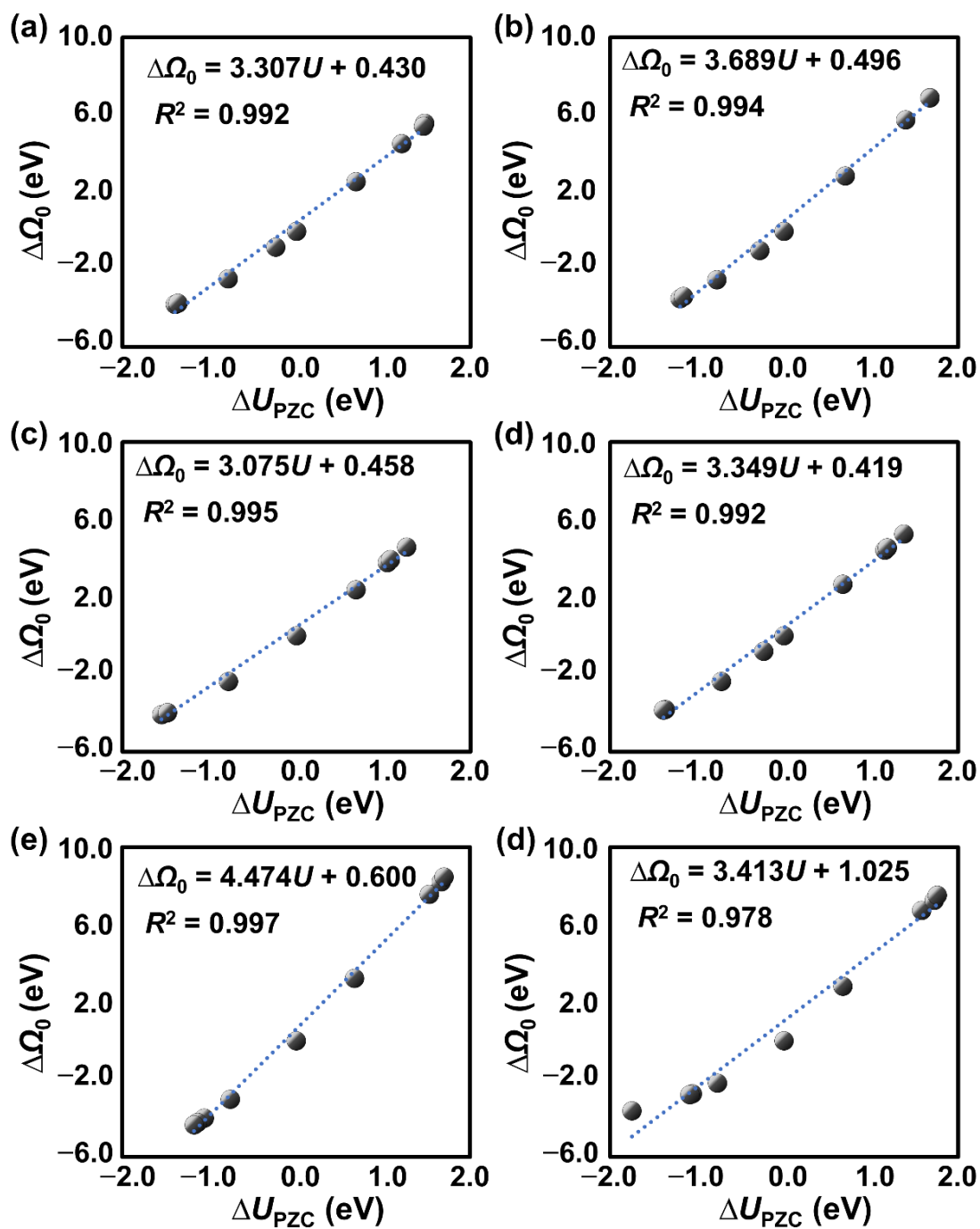

**Figure S6.** Grand canonical free energy relative to the zero charge free energy, $\Delta\Omega_0$ is plotted against the corresponding $\Delta U_{PZC}$, the system potential relative to the potential of zero charge (PZC), for (a) $R_a$, (b) $T_a$, (c) $R_{c1}$, (d) $T_{c1}$, (e) $R_{c2}$, and (f) $T_{c2}$ as shown in Fig. S2. The expression for linear regression and the $R^2$ value are depicted for each state.

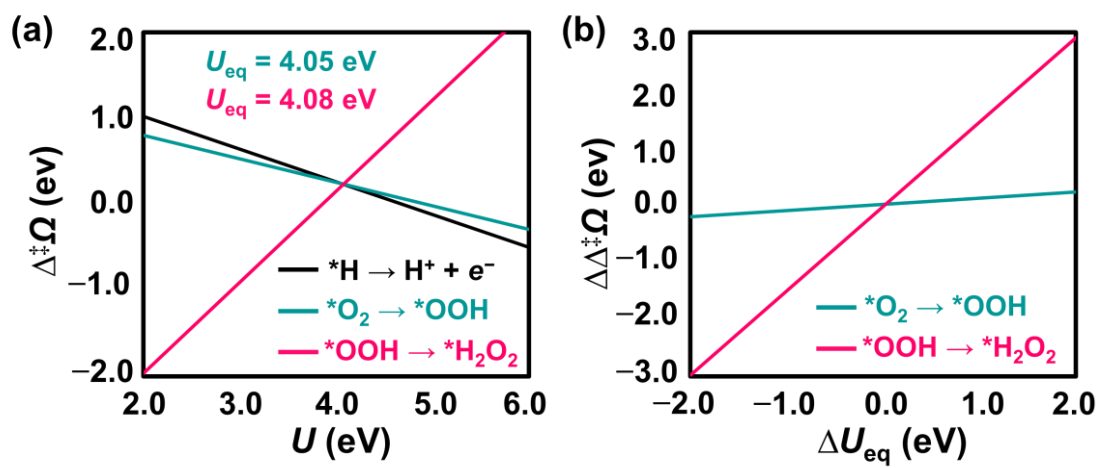

**Figure S7.** (a) Grand canonical free energy barrier, $\Delta^{\ddagger}\Omega$, plotted against the system potential $U$ for the anodic and cathodic reactions of the first and second protonation steps. The corresponding equilibrium potential, $U_{eq}$, is depicted. (b) Grand canonical free energy barrier difference between the anodic and cathodic reaction, $\Delta\Delta^{\ddagger}\Omega$, plotted against the corresponding $\Delta U_{eq}$, the system potential relative to $U_{eq}$ for the first and second protonation steps.

**Table S1.** Free energy decomposition of *$H_2O_2$ and $H_2O_{2(MA)}$

|  | $\Delta G$ | $\Delta E$ | $\Delta H_{corr}$ | $-TS$ | $\Delta G_{solv}$ |
|---|---|---|---|---|---|
| *$H_2O_2$ | -1.02 | -1.86 | 0.82 | −0.24 | −0.07 |
| $H_2O_{2(MA)}$ | -1.11 | -1.35 | 0.80 | −0.72 | −0.31 |